\documentclass[11pt,a4paper]{article}
\pdfoutput=1
\usepackage{amsmath,amssymb,tikz}
\usepackage{graphicx}
 \allowdisplaybreaks
\def\be{\begin{equation}}
\def\ee{\end{equation}}
\def\ba#1\ea{\begin{align}#1\end{align}}
\def\bg#1\eg{\begin{gather}#1\end{gather}}
\def\bm#1\em{\begin{multline}#1\end{multline}}
\def\bmd#1\emd{\begin{multlined}#1\end{multlined}}

\def\({\left(}
\def\){\right)}
\def\[{\left[}
\def\]{\right]}

\def \be {\begin{equation}}
\def \ee {\end{equation}}
\def \ba {\begin{array}}
\def \ea {\end{array}}
\def \bea{\begin{eqnarray}}
\def \eea{\end{eqnarray}}

\def\bea{\begin{eqnarray}}
\def\eea{\end{eqnarray}}

\newcommand{\bit}{\begin{itemize}}  \newcommand{\eit}{\end{itemize}}
\newcommand{\ben}{\begin{enumerate}}  \newcommand{\een}{\end{enumerate}}

\long\def\symbolfootnote[#1]#2{\begingroup%
\def\thefootnote{\fnsymbol{footnote}}\footnote[#1]{#2}\endgroup}
\newcommand{\sysu}{{\it School of Physics and Astronomy, Sun Yat-Sen University, 2 Daxue Road, Zhuhai 519082, China}}

\newcommand{\scnu}{{\it Guangdong Provincial Key Laboratory of Nanophotonic Functional Materials and Devices, South China Normal University, Guangzhou, 510631, China}}

\begin{document}
%%%%%%%%%%%%%%%%%%%%%%%%%%%%%%%%%%%%%%%%%%%
\thispagestyle{empty}
%%%%%%%%%%%%%%%%%%%%%%%%%%%%%%%%%%%%%%%%%%%
%%%%%%%%%%%%%%%%%%%%%%%%%%%%%%%%%%%%%%%%%%%
%\begin{flushright}
%\hfill{AEI-2015-xxx}
%\hfill{ NCTS-TH/1702}
%\end{flushright}
%%%%%%%%%%%%%%%%%%%%%%%%%%%%%%%%%%%%%%%%%%%
\begin{center}

~\vspace{20pt}

{\Large\bf Casimir Effect for Quantum Field theory in Networks}

\vspace{25pt}

Tian-Ming Zhao ${}^{1}$ \symbolfootnote[1]{Email:~\sf zhaotm@scnu.edu.cn}, Rong-Xin Miao ${}^{2}$\symbolfootnote[2]{Email:~\sf miaorx@mail.sysu.edu.cn}

\vspace{10pt}${{}^1}$\scnu

\vspace{10pt}${{}^2}$\sysu

\vspace{2cm}

\begin{abstract}

This paper studies quantum field theories defined in networks, which are the multi-branch generalizations of interface conformal field theory (ICFT). We propose a novel junction condition on the node and show that it is consistent with energy conservation in the sense that the total energy flow into the node is zero. As an application, we explore the Casimir effect on networks. Remarkably, the Casimir force on one edge can be changed from attractive to repulsive by adjusting the lengths of the other edges, providing a straightforward way to control the Casimir effect. We begin by discussing the Casimir effect for $(1+1)$-dimensional free massless scalars on a simple network. We then extend this discussion to various types of networks and higher dimensions. Finally, we offer brief comments on some open questions. 

\end{abstract}

\end{center}

%%%%%%%%%%%%%%%%%%%%%%%%%%%%%%%
\newpage
\setcounter{footnote}{0}
\setcounter{page}{1}
%%%%%%%%%%%%%%%%%%%%%%%%%%%%%%%

\tableofcontents
%%%%%%%%%%%%%%%%%%%%%%%%%%%%%%%

\section{Introduction}

Everything in the physical world is interconnected. Regardless of the type of particle—photons, electrons, protons—there exists universal gravitation between them. No matter how far apart the Einstein-Podolsky-Rosen pairs are separated, quantum entanglement persists between them \cite{Einstein:1935rr}. The concept of networks is an excellent tool for understanding complex connections and has significantly advanced artificial intelligence through the development of neural networks \cite{Hopfield:1982pe, Rumelhart:1986gxv, Hinton:2006tev}. Additionally, specific physical systems, such as chips and optical paths, naturally exhibit network structures. As a result, studying quantum field theories within networks is a crucial area of research that holds both theoretical and practical significance. This paper will begin by exploring the Casimir effect in the networks. 

\begin{figure}[t]
\centering
\includegraphics[width=9cm]{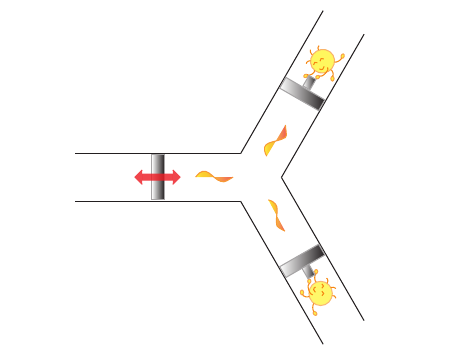}  
\caption{The Casimir force at the left edge can switch from attractive to repulsive by adjusting the lengths of the right edges.}
\label{NetCasimir}
\end{figure}

Casimir effect originates from changes in vacuum zero-point energy induced by boundaries or spacetime topology \cite{Casimir:1948dh, Plunien:1986ca, Bordag:2001qi,Milton:2004ya,Bordag:2009zz}. It has been experimentally measured \cite{Mohideen:1998iz, Bressi:2002fr, Klimchitskaya:2009cw} and demonstrates potential applications in nanotechnology. The Casimir effect can cause nanodevices to attract each other, potentially damaging precise structures or restricting their movement \cite{Chan:2001zza,Chan:2001zzb}. Thus, generating Casimir forces with controllable magnitude—particularly repulsive ones—holds significant importance in practical applications. There are several ways to produce repulsive Casimir force, such as imposing mixed boundary condition \cite{Bordag:2001qi}, driving the system out of equilibrium \cite{Antezza:2006bgd,Antezza:2005ngw}, and applying external fields \cite{Dean:2016atg,Du:2024gpi}. However, adjusting boundary conditions is impractical, and the other two methods only work in non-equilibrium states. In this article, we propose an alternative method for conveniently controlling the strength and direction of the Casimir force. We demonstrate that by changing the lengths of certain edges in a network, we can turn the Casimir force from attractive to repulsive at one edge. Refer to Fig. \ref{NetCasimir} for a schematic illustration.  

\begin{figure}[t]
\centering
\includegraphics[width=6cm]{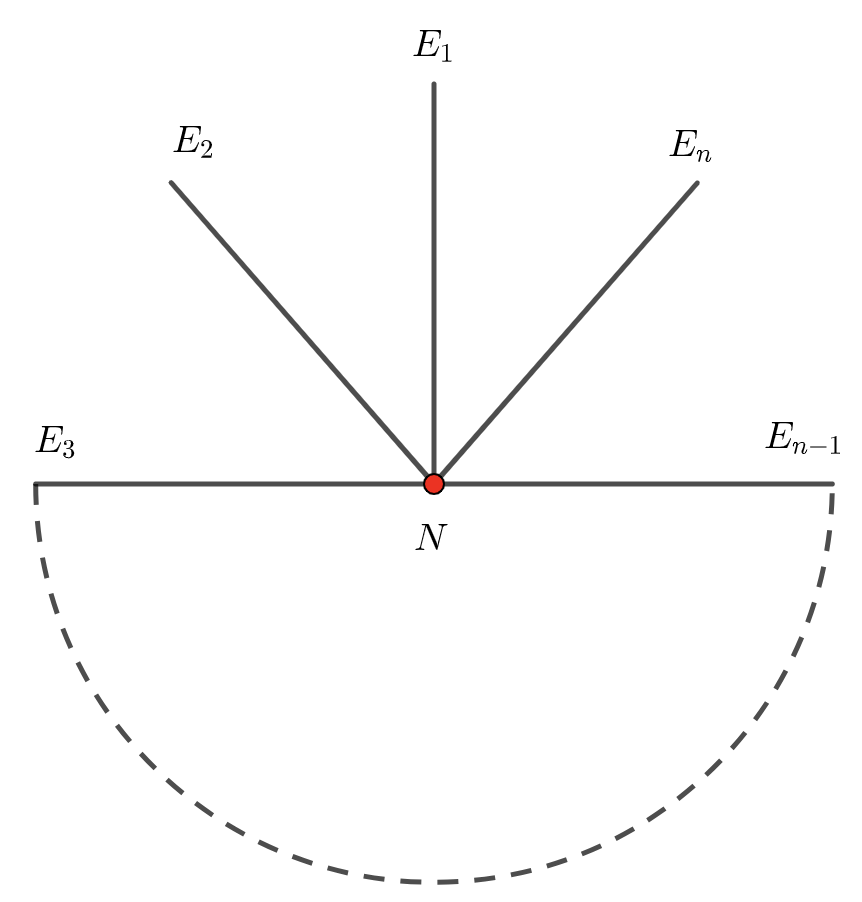}  
\caption{A network with $n$ edges $E_i$ linked by one node $N$ (redpoint).}
\label{network}
\end{figure}

Let’s summarize our key results. We derive the junction condition for quantum field theory at the network node $N$ and confirm it ensures the conservation of current, expressed as:
\begin{eqnarray} \label{current conservation}
\sum_{i=1}^n J_i|_{N}=0,
\end{eqnarray} 
where $J_i$ represents the current flowing from edge $E_i$ to node $N$, encompassing all types of current, such as energy and electric current. Refer to Fig. \ref{network} for the network containing $n$ edges connected by a single node. We first examine the Casimir effect for a $(1+1)$-dimensional massless scalar field on the simplest network with three edges connected at a single node. We impose the junction condition at the node and either a Neumann boundary condition (NBC) or a Dirichlet boundary condition (DBC) at the outer boundaries. Our results show that, for both boundary conditions, the Casimir force on one edge decreases as the lengths of the other edges increase. Specifically, for DBC, the Casimir force is always attractive, while for NBC, it can turn repulsive depending on the lengths of the edges. We explore more complex networks with additional nodes and loops, finding similar results. Finally, we extend our analysis to higher dimensions, where we again observe a repulsive Casimir force for NBC, but the force no longer changes signs. Instead, it approaches a non-zero limit when the lengths of the other edges become infinity.

The paper is organized as follows. In section 2, we formulate the quantum field theory on the network and derive the junction condition on the node. Section 3 examines the Casimir effect on the simplest network, which consists of three edges connected by a single node. We focus on the $(1+1)$-dimensional free massless scalar with various boundary conditions on the outer boundaries of networks. Section 4 and Section 5 generalize the discussions to other types of networks and higher dimensions, respectively. Finally, we conclude with some discussions in Section 6. The appendix discusses the network Casimir effect by applying the Green-function method. 

Notations: We take the natural units with $c=\hbar=1$ in this paper.

\section{Quantum field theories on networks}

The key aspect of developing a field theory on a network is determining the appropriate junction condition on the nodes. In this section, we apply the variational principle to derive this condition and confirm that it is consistent with the current conservation equation  (\ref{current conservation}). To demonstrate these ideas, we will focus on a massless free scalar field in the simplest network within $1+1$ dimensions. The generalizations to more general field theories, networks, and higher dimensions are straightforward.

Consider a network of $n$ edges $E$ connected at a single node $N$ in a $(1+1)$-dimensional flat spacetime. See Fig. \ref{network}. The action of a massless free scalar field is given by
\begin{eqnarray} \label{sect2: action}
I=\sum_{i=1}^n \int_{E_i} dt dx_i \Big( -\frac{1}{2} \eta^{\mu\nu} \partial_{\mu}\phi_i  \partial_{\nu}\phi_i  \Big),
\end{eqnarray} 
where $\eta^{\mu\nu}=\text{diag(-1,1)}$ is the metric, $\phi_i$ labels the scalar field on the edge $E_i$, and the edges $E_i$ intersect at the node $N$ with $x_i=0$.  Note that $x_i$ is the one-dimensional space coordinate on the edge $E_i$. Please don't confuse it with three-dimensional space coordinates, i.e., $x_i=(x,y,z)$. By varying the action, we derive the equation of motion (EOM) in the bulk
\begin{eqnarray} \label{sect2: EOM}
(\partial_t^2-\partial_{x_i}^2)\phi_i=0, 
\end{eqnarray} 
and the boundary terms on the node
\begin{eqnarray} \label{sect2: action1}
\delta I|_N= \int_{N} dt  \Big( \sum_{i=1}^n \partial_{n_i}\phi_i  \delta \phi_i \Big)=0,
\end{eqnarray}  
where $ \partial_{n_i}= \partial_{x_i}$ and $n_i$ is the normal vector pointing from node $N$ to edge $E_i$. For a well-defined variational principle, we need $\delta I|_N$ to vanish. Naturally, we also require continuity of the scalar fields at the node
\begin{eqnarray} \label{sect2: BC1a}
\phi_i|_N=\phi(t),
\end{eqnarray}  
for some arbitrary function $\phi(t)$. Then (\ref{sect2: action1}) becomes 
\begin{eqnarray} \label{sect2: action2}
\delta I|_N= \int_{N} dt  \Big(  \sum_{i=1}^n \partial_{n_i}\phi_i   \Big)\delta \phi =0,
\end{eqnarray}  
which yields 
\begin{eqnarray} \label{sect2: BC1b}
\sum_{i=1}^n \partial_{n_i}\phi_i |_N=0.
\end{eqnarray}  
In summary, we propose the following junction conditions at the node
\begin{eqnarray} \label{sect2: junction condition}
\text{junction conditions}: \ \phi_i|_N=\phi_j|_N,\  \ \sum_{i=1}^n \partial_{n_i}\phi_i |_N=0.
\end{eqnarray}

Two important points need to be addressed. First, the junction condition in equation (\ref{sect2: junction condition}) provides $n$ independent constraints, which is the correct number needed to define a theory on $n$ edges. To clarify, for a half-line with one boundary, one boundary condition is sufficient. Therefore, for a network with $n$ infinite edges, $n$ independent conditions are necessary. We will confirm that the junction condition is adequate for determining the Casimir effect on networks in the upcoming sections. Second, the junction condition is consistent with energy conservation at the nodes, as expressed in equation (\ref{current conservation}). The energy flow density for a scalar field is given by $J = T_{xt} = \partial_t \phi \partial_x \phi$. Using equations (\ref{sect2: BC1a}) and (\ref{sect2: BC1b}), we can show that the total energy flowing out of a node is zero
\begin{eqnarray} \label{sect2: conservation 1}
\sum_{i=1}^n J_i|_N=\sum_{i=1}^n \partial_t \phi_i \partial_{x_i} \phi_i |_N= (\sum_{i=1}^n  \partial_{n_i} \phi_i) \partial_t \phi |_N=0. 
\end{eqnarray}  

\begin{figure}[t]
\centering
\includegraphics[width=6cm]{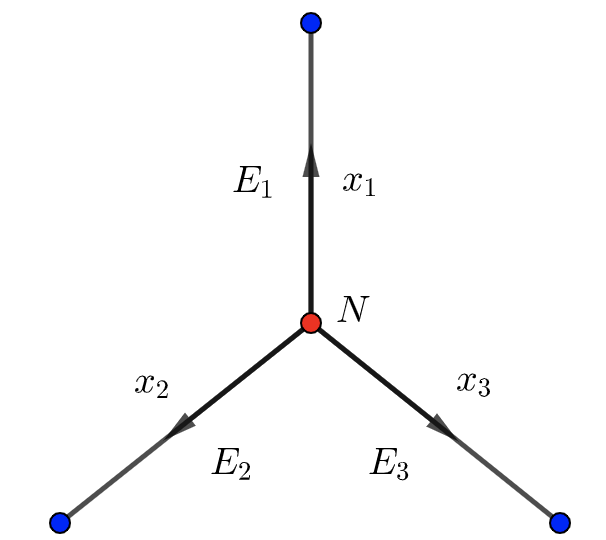} \includegraphics[width=6cm]{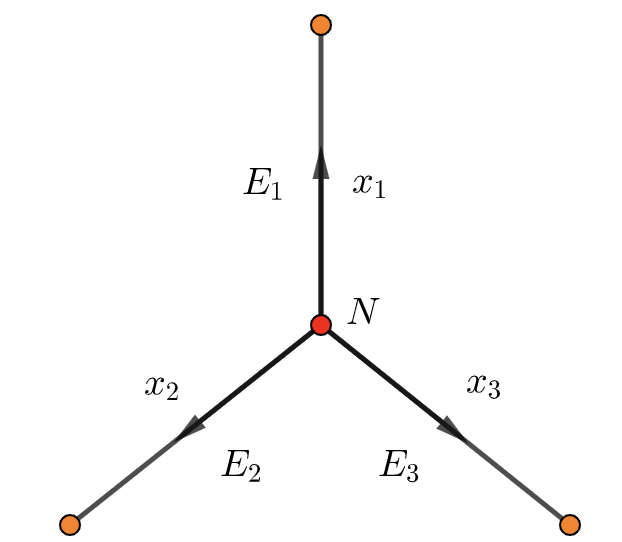} 
\caption{ Simplest network: three edges linked by one node. We impose the junction condition (\ref{sect2: junction condition}) on the node (red point), while Dirichlet and Neumann boundary conditions on the blue and yellow endpoints of the edges in the left and right figures, respectively.  Here $x_i$ is the space coordinate on edge $E_i$, which vanishes on the node (red point) $N: x_i=0$. The length of edge $E_i$ is $L_i$. }
\label{CrossLine}
\end{figure}

\section{Casimir effect on simplest networks}

This section investigates the Casimir effect on the simplest network with three edges connected by one node, as shown in Fig. \ref{CrossLine}. We impose the junction condition (\ref{sect2: junction condition}) on the node (redpoint), while Dirichlet boundary condition (DBC) and Neumann boundary condition (NBC) on the blue and yellow outer endpoints in the left and right figures, respectively. We find the Casimir effect is weaker than that of a strip for both DBC and NBC. With DBC, the Casimir energy is consistently negative, indicating an attractive force. In contrast, for NBC, the Casimir energy can switch signs by adjusting the edge lengths, allowing for control over the Casimir effect and enabling a shift from attraction to repulsion.

\subsection{DBC}

We first consider DBC on the outer boundaries $B$ of networks
\begin{eqnarray} \label{sect3: DBC}
{\text{Dirichlet BC}}:\ \phi_i |_B=0,
\end{eqnarray}  
where $B$ is labeled by blue points in Fig. \ref{CrossLine}. 

Solving the EOM (\ref{sect2: EOM}), we get the general solution for the scalar on edge $E_i$
\begin{eqnarray} \label{sect3: general solution}
\phi_i=\Big( a_i \sin(k x_i)+b_i \cos(k x_i) \Big) e^{-i \omega t},
\end{eqnarray}  
where $k= \omega$. To obey the continuous condition (\ref{sect2: BC1a}), the frequency $\omega$ on different edges must take the same value. Imposing DBC (\ref{sect3: DBC}) on the endpoint $x_i=L_i$, we can simplify (\ref{sect3: general solution}) to be
\begin{eqnarray} \label{sect3: solution1}
\phi_i= a_i \sin\Big(k(x_i-L_i)\Big) e^{-i \omega t},
\end{eqnarray} 
 where $L_i$ is the length of edge $E_i$. Substitute (\ref{sect3: solution1}) into the junction condition (\ref{sect2: junction condition}), we get 
 \begin{eqnarray} \label{sect3: constraint1}
a_1 \sin(k L_1)=a_2 \sin(k L_2)=a_3 \sin(k L_3),
\end{eqnarray} 
and
 \begin{eqnarray} \label{sect3: constraint2}
a_1 \cos(k L_1)+a_2 \cos(k L_2)+a_3 \cos(k L_3)=0. 
\end{eqnarray}  
Dividing (\ref{sect3: constraint2}) by (\ref{sect3: constraint1}) gives the constraint on the spectrum 
 \begin{eqnarray} \label{sect3: constraint}
\Delta(k)=\cot(k L_1)+\cot(k L_2)+\cot(k L_3)=0. 
\end{eqnarray} 

The following outlines the standard method for deriving the Casimir effect using the energy spectrum. We will use a straightforward mode-summing approach and will discuss the Green-function method in the appendix. The key formula we use is \cite{Bordag:2001qi}
\begin{eqnarray} \label{sect3: key trick}
\sum_n \frac{\omega_n}{2}-\sum_m \frac{p_m}{2}=\frac{1}{2\pi i}\oint \frac{\omega}{2} d \ln \Delta(\omega),
\end{eqnarray} 
where $\omega_n$ and $p_m$ are the zeros and poles of the spectrum function $\Delta(\omega)$ (\ref{sect3: constraint}), respectively. In our case, the poles read
\begin{eqnarray} \label{sect3: poles}
p_m= \frac{m \pi}{L_1}, \frac{m \pi}{L_2}, \frac{m \pi}{L_3},
\end{eqnarray} 
with $m$ being positive integers. Then, we get 
\begin{eqnarray} \label{sect3: poles sum}
\sum_m \frac{p_m}{2}=\frac{\pi}{2} (\frac{1}{L_1}+\frac{1}{L_2}+\frac{1}{L_3}) \sum_{m=1}^{\infty} m=-\frac{\pi}{24} (\frac{1}{L_1}+\frac{1}{L_2}+\frac{1}{L_3}),
\end{eqnarray} 
where we have used $\zeta(-1)=\sum m=-1/12$.  The contour integral of (\ref{sect3: key trick}) is performed along the right semi-circle with an infinite radius and the imaginary axis. As in the strip Casimir effect scenario, the integral around the infinite semi-circle vanishes with appropriate regularizations, leaving only the contribution from the imaginary axis. Thus, we have
\begin{eqnarray} \label{sect3: contour integral}
&&\frac{1}{2\pi i}\oint \frac{\omega}{2} d \ln \Delta(w)=\int_0^{\infty} -\frac{i \omega \Delta '(i \omega)}{2 \pi  \Delta (i \omega)} d\omega\nonumber\\
&&=\int_0^{\infty} \frac{\omega  \left(L_1 \text{csch}^2\left(L_1 \omega \right)+L_2 \text{csch}^2\left(L_2 \omega \right)+L_3 \text{csch}^2\left(L_3 \omega \right)\right)}{2 \pi  \left(\coth \left(L_1 \omega \right)+\coth \left(L_2 \omega \right)+\coth \left(L_3 \omega \right)\right)} d\omega,
\end{eqnarray} 
where the integral converges. From (\ref{sect3: key trick},\ref{sect3: poles sum},\ref{sect3: contour integral}), we finally obtain the Casimir energy for the simplest network with DBC endpoints (left figure of Fig. \ref{CrossLine}),
\begin{eqnarray} \label{sect3: Casimir energy DBC}
W_{\text{DBC}}&=&\sum_n \frac{\omega_n}{2}=-\frac{\pi}{24} (\frac{1}{L_1}+\frac{1}{L_2}+\frac{1}{L_3})\nonumber\\
&+&\int_0^{\infty} \frac{\omega  \left(L_1 \text{csch}^2\left(L_1 \omega \right)+L_2 \text{csch}^2\left(L_2 \omega \right)+L_3 \text{csch}^2\left(L_3 \omega \right)\right)}{2 \pi  \left(\coth \left(L_1 \omega \right)+\coth \left(L_2 \omega \right)+\coth \left(L_3 \omega \right)\right)} d\omega.
\end{eqnarray} 

Let us test the above formula in the symmetric case where $L_1=L_2=L_3=L$. In this scenario, the Casimir energy (\ref{sect3: Casimir energy DBC}) becomes $W_{DBC}=-\pi /(16 L)$. On the other hand, the spectrum can be obtained exactly for $ L_i = L$. From the constraints  (\ref{sect3: constraint1}, \ref{sect3: constraint2}), we obtain three independent solutions: one from $\cos(kL) = 0$ with $a_1 = a_2 = a_3$, and two from $\sin(kL) = 0$ resulting in $a_1 + a_2 + a_3 = 0$. These solutions contribute to the Casimir energy
\begin{eqnarray} \label{sect3: Casimir energy DBC L}
W_{\text{DBC}}|_{L_i=L}&=&  \sum_{m=0}^{\infty} \frac{\pi}{2L}(\frac{1}{2}+m) +2\sum_{n=1}^{\infty} \frac{n \pi}{2L} \nonumber\\
&=&\frac{\pi}{2L} \zeta \left(-1,\frac{1}{2}\right)+2 \frac{\pi}{2L}\zeta(-1)\\
&=&-\frac{\pi }{16 L},
\end{eqnarray} 
which agrees with (\ref{sect3: Casimir energy DBC}). Above, we have used $ \zeta \left(-1,\frac{1}{2}\right)=\sum_{m=0}^{\infty} (\frac{1}{2}+m)=1/24$. 

Let's look at a different scenario with $L_2 = L_3 \to \infty$. In this case, the Casimir energy (\ref{sect3: Casimir energy DBC}) becomes
\begin{eqnarray} \label{sect3: Casimir energy large L2L3}
\lim_{L_2=L_3\to \infty}W_{\text{DBC}}=-\frac{\text{Li}_2\left(\frac{1}{3}\right)}{4 \pi  L_1}\approx -\frac{0.029}{L_1} >-\frac{\pi}{24 L_1}.
\end{eqnarray} 
This result shows that the Casimir energy does not vanish, even when the network length is infinite. Additionally, the magnitude of this energy is smaller than that of a strip. These observations can be understood naturally. As waves on edge $E_1$ approach the node, they undergo reflection and transmission. Consequently, edge $E_1$ acts like a semi-transmissive strip, making the Casimir effect weaker but nonvanishing.

Now let us discuss the Casimir force
\begin{eqnarray} \label{sect3: Casimir force}
F^i=-\frac{\partial W}{\partial L_i},
\end{eqnarray} 
which is the experimentally measurable physical quantity. By integration by parts, we rewrite the Casimir energy (\ref{sect3: Casimir energy DBC}) as
\begin{eqnarray} \label{sect3: Casimir energy DBC parts}
&&W_{\text{DBC}}=\frac{-\pi}{24} (\frac{1}{L_1}+\frac{1}{L_2}+\frac{1}{L_3})
-\int_0^{\infty} \frac{\omega}{2\pi} d \log \left(\frac{\coth \left(L_1 \omega \right)+\coth \left(L_2 \omega \right)+\coth \left(L_3 \omega \right)}{3}\right)\nonumber\\
&&\ \ \ \ \ =\frac{-\pi}{24} (\frac{1}{L_1}+\frac{1}{L_2}+\frac{1}{L_3})+\int_0^{\infty}  \log\left( \frac{\coth \left(L_1 \omega \right)+\coth \left(L_2 \omega \right)+\coth \left(L_3 \omega \right)}{3}\right) \frac{d\omega}{2\pi}.
\end{eqnarray} 
We include the factor of 3 in the logarithm to ensure the integral converges as $\omega$ approaches infinity. Substituting this expression into the formula for the Casimir force yields
\begin{eqnarray} \label{sect3: Casimir force DBC}
F^i_{\text{DBC}}=-\frac{\pi}{24 L_i^2}+\int_0^{\infty} \frac{\omega  \text{csch}^2\left(L_i \omega \right)}{2 \pi  \left(\coth \left(L_1 \omega \right)+\coth \left(L_2 \omega \right)+\coth \left(L_3 \omega \right)\right)}d\omega.
 \end{eqnarray} 
Comparing the Casimir energy (\ref{sect3: Casimir energy DBC}) with the force (\ref{sect3: Casimir force DBC}), we observe an interesting relation 
\begin{eqnarray} \label{sect3: Casimir relation}
W=F^1 L_1+F^2 L_2+F^3 L_3. 
 \end{eqnarray} 
Let us explain the physical origin of this novel relation. In a $(1+1)$-dimensional conformal field theory (CFT), the stress-energy tensor must be traceless $T^{\mu}_{\ \mu}=-T_{tt}+T_{xx}=0$, resulting in equal energy density $T_{tt}$ and pressure $T_{xx}=F$ along the edges. Thus, the total energy can be expressed as the sum of forces multiplied by their respective lengths.

Let us discuss some typical cases. For $L_i=L$, the force (\ref{sect3: Casimir force DBC}) can be calculated exactly. We get
\begin{eqnarray} \label{sect3: Casimir force DBC L}
F^i_{\text{DBC}}|_{L_i=L}=-\frac{\pi }{48 L^2},
 \end{eqnarray} 
which matches the Casimir energy (\ref{sect3: Casimir energy DBC L}) since we have three edges involved. As we let $L_2 = L_3 \to \infty$, the force becomes 
\begin{eqnarray} \label{sect3: Casimir force DBC large L2L3}
\lim_{L_2=L_3\to \infty}F^1_{\text{DBC}}=-\frac{\text{Li}_2\left(\frac{1}{3}\right)}{4 \pi  L_1^2}, 
 \end{eqnarray} 
which agrees with (\ref{sect3: Casimir energy large L2L3}). Now, if we set $L_2 = L_3 = 0$, the integral from equation (\ref{sect3: Casimir force DBC}) becomes zero, and the force on edge $E_1$ simplifies to that of a strip
\begin{eqnarray} \label{sect3: Casimir force DBC small L2L3}
\lim_{L_2=L_3\to 0}F^1_{\text{DBC}}=-\frac{\pi}{24 L_1^2}. 
 \end{eqnarray} 
This result is expected because the network becomes a strip when $L_2 = L_3 = 0$.

\subsection{NBC}

Let's go on to explore the network with NBC
\begin{eqnarray} \label{sect3: NBC}
{\text{Neumann BC}}:\ \partial_{x_i}\phi_i |_B=0,
\end{eqnarray}  
where $B$  refers to the outer endpoints of the edges (marked as yellow points in Fig. \ref{CrossLine}). The calculations for NBC follow a similar process as those for DBC. Below are the key points and main results for NBC.

To satisfy NBC at the outer endpoints, we make the following ansatz of scalars 
\begin{eqnarray} \label{sect3: solution NBC}
\phi_i= a_i \cos\Big(k(x_i-L_i)\Big) e^{-i \omega t}.
\end{eqnarray} 
Then the junction condition (\ref{sect2: junction condition}) on the node gives
 \begin{eqnarray} \label{sect3: constraint1 NBC}
a_1 \cos(k L_1)=a_2 \cos(k L_2)=a_3 \cos(k L_3),
\end{eqnarray} 
and
 \begin{eqnarray} \label{sect3: constraint2 NBC}
a_1 \sin(k L_1)+a_2 \sin(k L_2)+a_3 \sin(k L_3)=0. 
\end{eqnarray}  
From the above two equations, we get the spectrum for NBC
 \begin{eqnarray} \label{sect3: constraint NBC}
\Delta(k)=\tan(k L_1)+\tan(k L_2)+\tan(k L_3)=0,
\end{eqnarray} 
which is similar to the spectrum for DBC, except that we replace $\cot$ with $\tan$. Therefore, when calculating the Casimir effect, we substitute $\text{coth}$ and $\text{csch}$ with $\tanh$ and $\text{sech}$. Additionally, the poles of $\Delta(k)$ shift to
\begin{eqnarray} \label{sect3: poles NBC}
p_m= \frac{(\frac{1}{2}+m)\pi}{L_1}, \frac{(\frac{1}{2}+m) \pi}{L_2}, \frac{(\frac{1}{2}+m) \pi}{L_3}. 
\end{eqnarray} 
Taking these modifications into account and following the methodology in sect. 3.1, we obtain the Casimir energy
\begin{eqnarray} \label{sect3: Casimir energy NBC}
W_{\text{NBC}}&=&\frac{\pi }{48}  \left(\frac{1}{L_1}+\frac{1}{L_2}+\frac{1}{L_3}\right)\nonumber\\
&-&\int_0^{\infty} \frac{\omega  \left(L_1 \text{sech}^2\left(L_1 \omega \right)+L_2 \text{sech}^2\left(L_2 \omega \right)+L_3 \text{sech}^2\left(L_3 \omega \right)\right)}{2 \pi  \left(\tanh \left(L_1 \omega \right)+\tanh \left(L_2 \omega \right)+\tanh \left(L_3 \omega \right)\right)} d\omega,
\end{eqnarray} 
and the Casimir force for NBC
\begin{eqnarray} \label{sect3: Casimir force NBC}
F^i_{\text{NBC}}=\frac{\pi}{48 L_i^2}-\int_0^{\infty} \frac{\omega  \text{sech}^2\left(L_i \omega \right)}{2 \pi  \left(\tanh \left(L_1 \omega \right)+\tanh \left(L_2 \omega \right)+\tanh \left(L_3 \omega \right)\right)} d\omega.
 \end{eqnarray}
The above Casimir energy and force for NBC obey the conformal relation (\ref{sect3: Casimir relation}), reinforcing the validity of our findings.

To gain a deeper understanding of the Casimir effect for NBC, let us examine some typical examples. Surprisingly, the Casimir effect disappears for $L_i=L$
\begin{eqnarray} \label{sect3: Casimir force NBC L}
W_{\text{NBC}}|_{L_i=L}=F^i_{\text{NBC}}|_{L_i=L}=0.
 \end{eqnarray} 
We can confirm it through spectrum analysis. From the spectrum (\ref{sect3: constraint NBC}) with $L_i=L$, we get one independent solution with $\sin(k L)=0, a_1=a_2=a_3$, and other two independent solutions with $\cos(kL)=0, a_1+a_2+a_3=0$. Summing these modes gives us a zero Casimir effect for NBC
\begin{eqnarray} \label{sect3: Casimir energy NBC L 1}
W_{\text{NBC}}|_{L_i=L}&=&  2\sum_{m=0}^{\infty} \frac{\pi}{2L}(\frac{1}{2}+m) +\sum_{n=1}^{\infty} \frac{n \pi}{2L} \nonumber\\
&=&\frac{\pi}{L} \zeta \left(-1,\frac{1}{2}\right)+\frac{\pi}{2L}\zeta(-1)=0.
\end{eqnarray} 
Now, as $L_2 = L_3 \to \infty$, we find
\begin{eqnarray} \label{sect3: Casimir energy NBC large L2L3}
&&\lim_{L_2=L_3\to \infty}W_{\text{NBC}}=-\frac{\text{Li}_2\left(-\frac{1}{3}\right)}{4 \pi  L_1}\approx \frac{0.025}{L_1},  \\
&&\lim_{L_2=L_3\to \infty}F^1_{\text{NBC}}= -\frac{\text{Li}_2\left(-\frac{1}{3}\right)}{4 \pi  L_1^2}\approx \frac{0.025}{L_1^2}.  \label{sect3: Casimir force NBC large L2L3}
 \end{eqnarray} 
On the other hand, as $L_2 = L_3 \to 0$
 \begin{eqnarray} \label{sect3: Casimir energy NBC zero L2L3}
\lim_{L_2=L_3\to 0}F^1_{\text{NBC}}=-\frac{\pi}{24 L_1^2},
 \end{eqnarray} 
which aligns with the Casimir force for a strip. From our findings in the two limits, we see that the Casimir force can change signs under NBC.
 
 \begin{figure}[t]
\centering
\includegraphics[width=7cm]{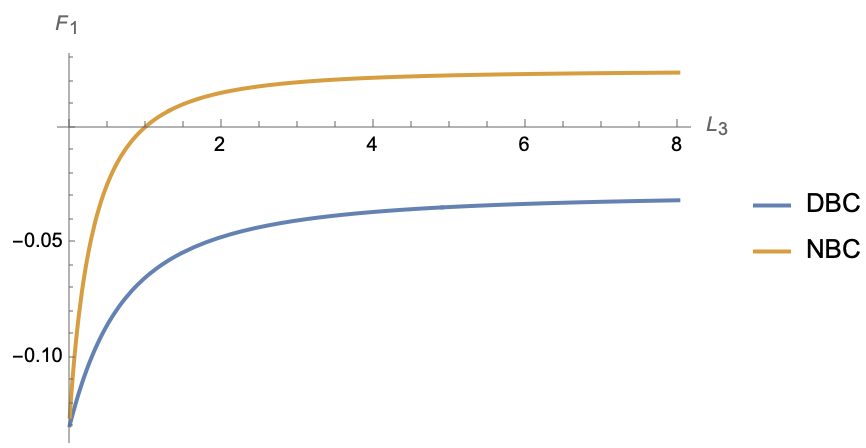} \includegraphics[width=7cm]{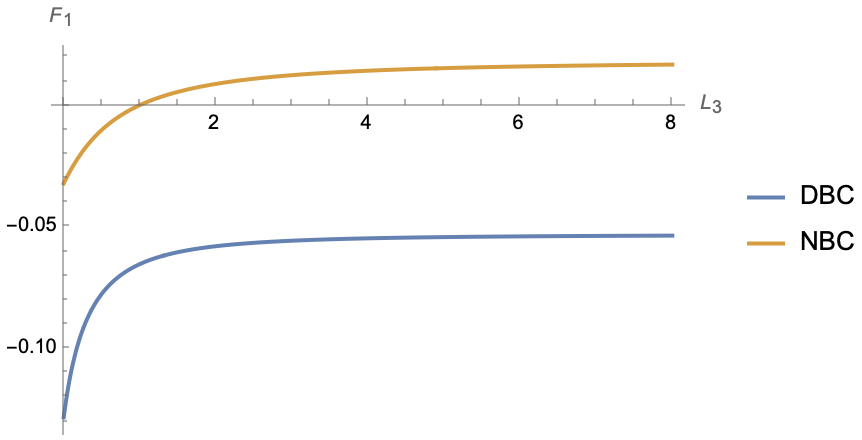} 
\caption{ Casimir force on edge $E_1$ varies with the lengths of other edges. The left figure is for $L_1=1, L_2=L_3$, and the right figure is for $L_1=L_2=1$. For DBC (blue curve), the amplitude of force $F_1$ decreases with $L_3$ but remains the same signs. While for NBC (orange curve), the force $F_1$ flips signs as $L_3$ increases. For both DBC and NBC, $F_1$ approaches to a constant value in the large $L_3$ limit.  }
\label{3lineF1}
\end{figure}

To conclude this section, we illustrate the Casimir force $F_1(L_3)$ in Fig. \ref{3lineF1}, where $L_3$ represents the length of edge $E_3$. The figure demonstrates that the amplitude of the force $F_1$ decreases as $L_3$ increases while maintaining the same sign for DBC. In contrast, for NBC, the force $F_1$ changes sign as $L_3$ increases. For both DBC and NBC, $F_1$ approaches a constant value in the limit of large $L_3$, which is consistent with the expressions in equations (\ref{sect3: Casimir force DBC large L2L3}) and (\ref{sect3: Casimir force NBC large L2L3}).

\section{Casimir effect on general networks}

This section outlines a method for deriving the Casimir effect on various types of networks. This approach is beneficial for programming. To illustrate the process, we'll use simple tree and loop networks as examples. Detailed calculations for more complex networks will be addressed in future work \cite{Tianming}.

 \begin{figure}[t]
\centering
\includegraphics[width=3.1cm]{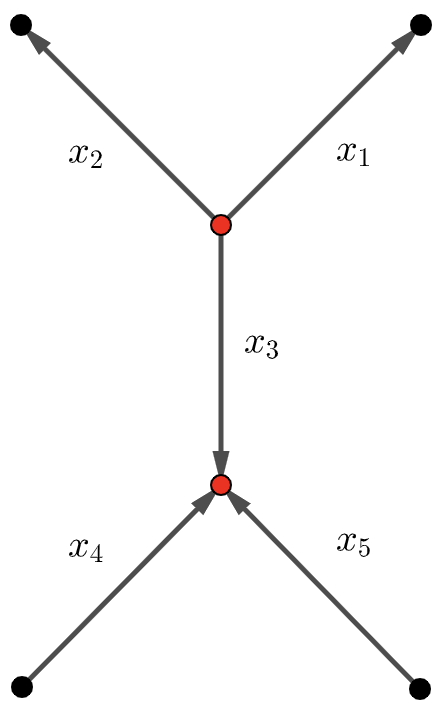}  \ \ \  \ \ \ \ \ \ \ \ \ \ \ \ \ \ \ \ \ \includegraphics[width=2.5cm]{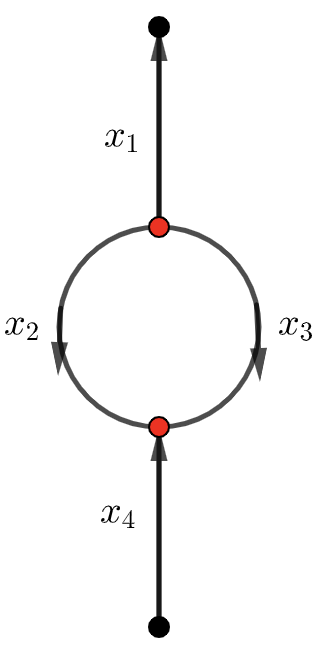}  
\caption{Tree (left) and loop (right) networks with two nodes.  Note that the arrows label the directions of the coordinates $0\le x_i\le L_i$. It does not mean it is a directed network. We impose the junction condition (\ref{sect2: junction condition}) on the nodes (red points), while either DBC (\ref{sect3: DBC}) or NBC (\ref{sect3: NBC}) on the outside edge endpoints (black points).}
\label{Tree}
\end{figure}

\subsection{Tree network}

This subsection examines the Casimir effect on a network with five edges linked by two nodes, as illustrated in Fig. \ref{Tree} (left). Note that the pointed tip's front segment corresponds to $x_i=L_i$, and the rear segment corresponds to $x_i=0$. We impose the junction condition (\ref{sect2: junction condition}) on the nodes (red points), while either the DBC (\ref{sect3: DBC}) or NBC (\ref{sect3: NBC}) is applied to the outer endpoints of the edges (black points). 

Let us first study the case of DBC. We make the following ansatzes for the scalars on each edge 
\begin{eqnarray} \label{sect4: scalar1}
&&\phi_i= a_i \sin\Big(k(x_i-L_i)\Big) e^{-i \omega t}, \ \text{for i=1,2,}\\
&&\phi_3=\Big(a_3 \sin(kx_3)+b_3 \cos(kx_3) \Big) e^{-i \omega t},  \label{sect4: scalar2} \\
&&\phi_j= a_j \sin(k x_j)e^{-i \omega t}, \ \text{for j=4,5},  \label{sect4: scalar3}
\end{eqnarray} 
where $0\le x_i\le L_i$. One can easily see that the above ansatzes obey DBC (\ref{sect3: DBC}) on the outer endpoints of Fig. \ref{Tree}. The junction condition (\ref{sect2: junction condition}) on the node $(x_1=x_2=x_3=0)$ yields
\begin{eqnarray} \label{sect4: node1a}
&& a_1 \sin(kL_1)-a_2 \sin(kL_2)=0,\\ \label{sect4: node1b}
&&a_2 \sin(kL_2)+b_3=0,\\ \label{sect4: node1c}
&& a_1 \cos(k L_1)+a_2 \cos(k L_2)+a_3=0.
\end{eqnarray} 
Similarly, the junction condition (\ref{sect2: junction condition}) on the node $(x_3=L_3, x_4=L_4, x_5=L_5)$ gives
\begin{eqnarray} \label{sect4: node2a}
&& \Big(a_3 \sin(kL_3)+b_3 \cos(kL_3) \Big)-a_4 \sin(k L_4)=0,\\ \label{sect4: node2b}
&&a_4 \sin(k L_4)-a_5 \sin(k L_5)=0,\\ \label{sect4: node2c}
&& a_3 \cos(k L_3)-b_3 \sin(k L_3)+a_4 \cos(k L_4)+a_5 \cos(k L_5)=0.
\end{eqnarray} 
We rewrite the above six equations into the matrix form $M\cdot a^T=0$, where $a=(a_1,a_2,a_3,b_3,a_4,a_5)$ and 
\begin{eqnarray} \label{sect4: matrix}
M=\left(
\begin{array}{cccccc}
 \sin \left(k L_1\right) & -\sin \left(k L_2\right) & 0 & 0 & 0 & 0 \\
 0 & \sin \left(k L_2\right) & 0 & 1 & 0 & 0 \\
 \cos \left(k L_1\right) & \cos \left(k L_2\right) & 1 & 0 & 0 & 0 \\
 0 & 0 & \sin \left(k L_3\right) & \cos \left(k L_3\right) & -\sin \left(k L_4\right) & 0 \\
 0 & 0 & 0 & 0 & \sin \left(k L_4\right) & -\sin \left(k L_5\right) \\
 0 & 0 & \cos \left(k L_3\right) & -\sin \left(k L_3\right) & \cos \left(k L_4\right) & \cos \left(k L_5\right) \\
\end{array}
\right).
\end{eqnarray} 
To have non-zero solutions to $M\cdot a^T=0$, we must have $|M|=0$, which gives the spectrum
\begin{eqnarray} \label{sect4: spectrum}
&&\Delta_{\text{DBC}}(k)=\sin \left(k L_1\right) \sin \left(k L_2\right) \sin \left(k L_4\right) \cos \left(k L_3\right) \cos \left(k L_5\right)\nonumber\\
&&+\sin \left(k L_5\right) \cos \left(k L_3\right) \left(\sin \left(k \left(L_1+L_2\right)\right) \sin \left(k L_4\right)+\sin \left(k L_1\right) \sin \left(k L_2\right) \cos \left(k L_4\right)\right)\\
&&+\sin \left(k L_3\right) \left(\sin \left(k \left(L_1+L_2\right)\right) \sin \left(k \left(L_4+L_5\right)\right)-\sin \left(k L_1\right) \sin \left(k L_2\right) \sin \left(k L_4\right) \sin \left(k L_5\right)\right)=0 .\nonumber
\end{eqnarray} 
Note that, unlike those of sect.3, the spectrum (\ref{sect4: spectrum}) obtained from $|M|=0$ has no poles. However, the cost is that the contour integral of (\ref{sect3: key trick}) is divergent 
\begin{eqnarray} \label{sect4: divergent}
W_0=\frac{1}{2\pi i}\oint \frac{\omega}{2} d \ln \Delta(w)=\int_0^{\infty} -\frac{i \omega \Delta '(i \omega)}{2 \pi  \Delta (i \omega)} d\omega \to \infty,
\end{eqnarray} 
and we need further regularizations. By taking the large frequency limit, we have
\begin{eqnarray} \label{sect4: large w}
\lim_{\omega\to \infty} -\frac{i \omega \Delta '(i \omega)}{2 \pi  \Delta (i \omega)}  \to-\frac{\omega \sum_{i=1}^5 L_i}{2\pi}.
\end{eqnarray} 
Subtracting the above divergence, we get the finite Casimir energy
\begin{eqnarray} \label{sect4: Casimir energy}
W=\int_0^{\infty}\Big( \frac{\omega \sum_{i=1}^n L_i}{2\pi}-\frac{i \omega \Delta '(i \omega)}{2 \pi  \Delta (i \omega)} \Big) d\omega,
\end{eqnarray} 
where $n=5$, $\Delta (k)$ is given by (\ref{sect4: spectrum}).  We remark that the method of this section yields the same results as those in sect.3 when applied to the cases in sect.3. $ F^i=-\partial W/\partial L_i$ can derive the Casimir force. We have checked numerous examples and verified the conformal relation $W=\sum_{i=1}^5 F^i L_i$ numerically. It is a non-trivial test of our calculations. When $L_i=L$, (\ref{sect4: Casimir energy}) becomes 
\begin{eqnarray} \label{sect4: Casimir energy DBC}
W_{\text{DBC}}|_{L_i=L}=\int_0^{\infty} \frac{L \omega  (1-\coth (L \omega )) (-16 \sinh (2 L \omega )+29 \cosh (2 L \omega )+19)}{2 \pi  (9 \cosh (2 L \omega )+7)}d\omega\approx -\frac{0.280}{L}.
\end{eqnarray} 

The case for NBC (\ref{sect3: NBC}) is similar. The scalars obeying NBC (\ref{sect3: NBC}) reads
\begin{eqnarray} \label{sect4: scalar1NBC}
&&\phi_i= a_i \cos\Big(k(x_i-L_i)\Big) e^{-i \omega t}, \ \text{for i=1,2,}\\
&&\phi_3=\Big(a_3 \sin(kx_3)+b_3 \cos(kx_3) \Big) e^{-i \omega t},  \label{sect4: scalar2NBC} \\
&&\phi_j= a_j \cos(k x_j)e^{-i \omega t}, \ \text{for j=4,5}. \label{sect4: scalar3NBC}
\end{eqnarray} 
Following the same approach, we derive the spectrum for NBC
\begin{eqnarray} \label{sect4: spectrum NBC}
&&\Delta_{\text{NBC}}(k)=-\frac{1}{4} \left(\sin \left(k \left(L_1+L_3-L_4+L_5\right)\right)+3 \sin \left(k \left(L_1+L_3+L_4+L_5\right)\right)\right) \cos \left(k L_2\right)\nonumber\\
&&-\frac{1}{4} \left(\sin \left(k \left(L_1+L_3+L_4-L_5\right)\right)-\sin \left(k \left(L_1+L_3-L_4-L_5\right)\right)\right) \cos \left(k L_2\right)\\
&&+\sin \left(k L_2\right) \cos \left(k L_1\right) \left(\sin \left(k L_3\right) \sin \left(k \left(L_4+L_5\right)\right)-\cos \left(k L_3\right) \cos \left(k L_4\right) \cos \left(k L_5\right)\right)=0 .\nonumber
\end{eqnarray} 
Substituting (\ref{sect4: spectrum NBC}) into (\ref{sect4: Casimir energy}), we derive the Casimir energy for NBC. For the case $L_i=L$, we get a positive Casimir energy
\begin{eqnarray} \label{sect4: Casimir energy NBC}
W_{\text{NBC}}|_{L_i=L}=\int_0^{\infty} \frac{L w \left(-45 e^{-4 L w}+e^{-2 L w}+11 e^{2 L w}-7\right)}{2 \pi  (2 \sinh (2 L w)+9 \sinh (4 L w))} d\omega\approx \frac{0.046}{L},
\end{eqnarray} 
which implies a repulsive force.

\begin{figure}[t]
\centering
\includegraphics[width=10cm]{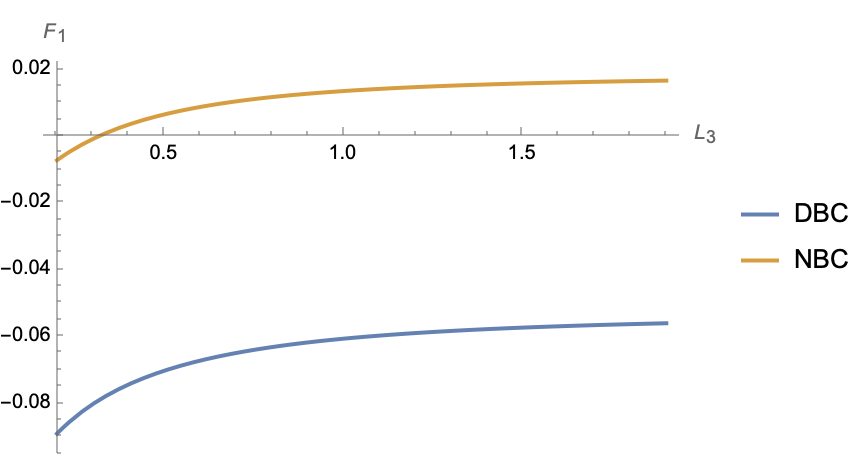}   
\caption{Casimir force for the tree network (Fig. \ref{Tree} (left)) with  $L_1=L_2$ and $L_3=L_4=L_5$. Interestingly, for the NBC, the forces $F_1=F_2$ on the outside edges $E_1, E_2$ change signs when increasing the lengths of other edges.  }
\label{forceII}
\end{figure}

To conclude this subsection, we analyze the Casimir force, denoted as $ F_1 = F_2$, acting on the edges $E_1$ and $E_2$ for the specific case where $L_1 = L_2$ and $L_3 = L_4 = L_5$, as illustrated in Fig. \ref{forceII}. Similar to the situation discussed in Section 3, we observe that for NBC, the forces $ F_1$ and $ F_2$ change signs when the lengths of the other edges are increased.

\subsection{Loop network}

This subsection examines the Casimir effect on a network with two nodes and one loop, as illustrated in Fig. \ref{Tree} (right). Let us start with the scalars obeying DBC
\begin{eqnarray} \label{sect5: scalar1}
&&\phi_1= a_1 \sin\Big(k(x_1-L_1)\Big) e^{-i \omega t},\\
&&\phi_j=\Big(a_j \sin(kx_j)+b_j \cos(kx_j) \Big) e^{-i \omega t},  \ \text{for j=2,3,}\label{sect5: scalar2} \\
&&\phi_4= a_4 \sin(kx_4) e^{-i \omega t} \label{sect5: scalar3} 
\end{eqnarray} 
The junction conditions on the nodes $(x_1=x_2=x_3=0)$ and $(x_2=L_2, x_3=L_3, x_4=L_4)$ yield six equations 
\begin{eqnarray} \label{sect5: node1a}
&& a_1 \sin(kL_1)+b_2=0\\ \label{sect5: node1b}
&&b_2-b_3=0,\\ \label{sect5: node1c}
&& a_1 \cos(k L_1)+a_2 +a_3=0,\\  \label{sect5: node2a}
&&a_2 \sin(k L_2)+b_2 \cos(k L_2)-a_4 \sin(k L_4)=0, \\  \label{sect5: node2b}
&&a_4 \sin(k L_4)-a_3 \sin(k L_3)-b_3 \cos(k L_3)=0,\\  \label{sect5: node2c}
&&a_2\cos(k L_2)-b_2\sin(k L_2)+a_3\cos(k L_3)-b_3\sin(k L_3)+a_4 \cos(k L_4)=0.
\end{eqnarray} 
The above equations can be rewritten into the matrix form $M\cdot a^T=0$ with $a=(a_1, a_2,b_2,a_3,b_3,a_4)$ and 
\begin{eqnarray} \label{sect5: metrix}
M=\left(
\begin{array}{cccccc}
 \sin \left(k L_1\right) & 0 & 1 & 0 & 0 & 0 \\
 0 & 0 & 1 & 0 & -1 & 0 \\
 \cos \left(k L_1\right) & 1 & 0 & 1 & 0 & 0 \\
 0 & \sin \left(k L_2\right) & \cos \left(k L_2\right) & 0 & 0 & -\sin \left(k L_4\right) \\
 0 & 0 & 0 & -\sin \left(k L_3\right) & -\cos \left(k L_3\right) & \sin \left(k L_4\right) \\
 0 & \cos \left(k L_2\right) & -\sin \left(k L_2\right) & \cos \left(k L_3\right) & -\sin \left(k L_3\right) & \cos \left(k L_4\right) \\
\end{array}
\right).
\end{eqnarray} 
To have non-zero solutions, we require the determinant of $M$ vanishes, yielding the spectrum 
\begin{eqnarray} \label{sect5: loop spectrum DBC}
&&\Delta_{\text{DBC}}(k)= -\cos \left(k L_1\right)\sin \left(k L_2\right) \sin \left(k L_3\right) \cos \left(k L_4\right)\nonumber\\
&&+4 \sin \left(k L_1\right) \sin ^2\left(\frac{1}{2} k \left(L_2+L_3\right)\right) \sin \left(k L_4\right)-\sin \left(k \left(L_2+L_3\right)\right) \sin \left(k \left(L_1+L_4\right)\right).
\end{eqnarray} 
Substituting the above spectrum into (\ref{sect4: Casimir energy}) with $n=4$, we obtain the renormalized Casimir energy for DBC. For simplicity, we do not show the complicated expressions. For the special case $L_i=L$, we have
\begin{eqnarray} \label{sect5: Casimir energy DBC L}
W_{\text{DBC}}|_{L_i=L}=\int_0^{\infty} \frac{L \omega  (1-\coth (L \omega )) (-5 \sinh (2 L \omega )+13 \cosh (2 L \omega )-3)}{9 \pi  \cosh (2 L \omega )+\pi } d\omega\approx-\frac{0.216}{L}.
\end{eqnarray} 

Now, let us go on to discuss the case of NBC. The scalars obeying NBC (\ref{sect3: NBC}) take the following form
\begin{eqnarray} \label{sect5: scalar1 NBC}
&&\phi_1= a_1 \cos\Big(k(x_1-L_1)\Big) e^{-i \omega t},\\
&&\phi_j=\Big(a_j \sin(kx_j)+b_j \cos(kx_j) \Big) e^{-i \omega t},  \ \text{for j=2,3,}\label{sect5: scalar2 NBC} \\
&&\phi_4= a_4 \cos(kx_4) e^{-i \omega t} \label{sect5: scalar3 NBC} .
\end{eqnarray} 
The junction condition (\ref{sect2: junction condition}) on nodes give the spectrum 
\begin{eqnarray} \label{sect5: loop spectrum NBC}
&&\Delta_{\text{NBC}}(k)=\sin \left(k L_1\right) \left(\sin \left(k L_2\right) \cos \left(k L_3\right) \cos \left(k L_4\right)+\sin \left(k L_3\right) \cos \left(k \left(L_2+L_4\right)\right)\right)\\
&&+\sin \left(\frac{1}{2} k \left(L_2+L_3\right)\right) \left(3 \sin \left(\frac{1}{2} k \left(L_2+L_3+2 L_4\right)\right)+\sin \left(\frac{1}{2} k \left(L_2+L_3-2 L_4\right)\right)\right) \cos \left(k L_1\right).\nonumber
\end{eqnarray} 
 Substituting the above spectrum into (\ref{sect4: Casimir energy}) with $n=4$, we obtain the renormalized Casimir energy for NBC. For  $L_i=L$, we get
 \begin{eqnarray} \label{sect5: Casimir energy NBC L}
W_{\text{NBC}}|_{L_i=L}=\int_0^{\infty} \frac{4 L \omega  (1-\coth (L \omega )) (\sinh (L \omega )-2 \cosh (L \omega ))^2}{\pi  (9 \cosh (2 L \omega )+7)} d\omega\approx-\frac{0.149}{L}.
\end{eqnarray} 
 
Let us test our results in two special cases. First, when $L_1 = L_4 = L/2$ and $L_2 = L_3 \to 0$, the loop disappears, simplifying the network (as shown in Fig. \ref{Tree} (right)) to a strip with width $L$. In this scenario, the network Casimir energy (\ref{sect4: Casimir energy}) matches that of a strip for both DBC and NBC
 \begin{eqnarray} \label{sect5: zero L2 L3}
\lim_{L_2=L_3\to 0}W=\int_0^{\infty } -\frac{\omega  \left(e^{2 L \omega } (2 L \omega -1)+1\right)}{\pi  \left(e^{2 L \omega }-1\right)^2} \, d\omega=-\frac{\pi }{24 L^2}.
\end{eqnarray} 
This provides strong support for our results. Next, when $L_1 = L_4 = 0$ and $L_2 = L_3 = L/2$, the network transforms into a circle with perimeter $L$, and the Casimir energy becomes for both DBC and NBC
 \begin{eqnarray} \label{sect5: zero L2 L3}
\lim_{L_1=L_4\to 0}W=\int_0^{\infty } -\frac{\omega  \left(e^{L \omega } (L \omega -1)+1\right)}{\pi  \left(e^{L \omega }-1\right)^2} \, d\omega=-\frac{\pi }{6 L^2},
\end{eqnarray} 
which agrees with the circle Casimir energy \cite{Bordag:2001qi}. It reinforces our results once again.

\begin{figure}[t]
\centering
\includegraphics[width=7cm]{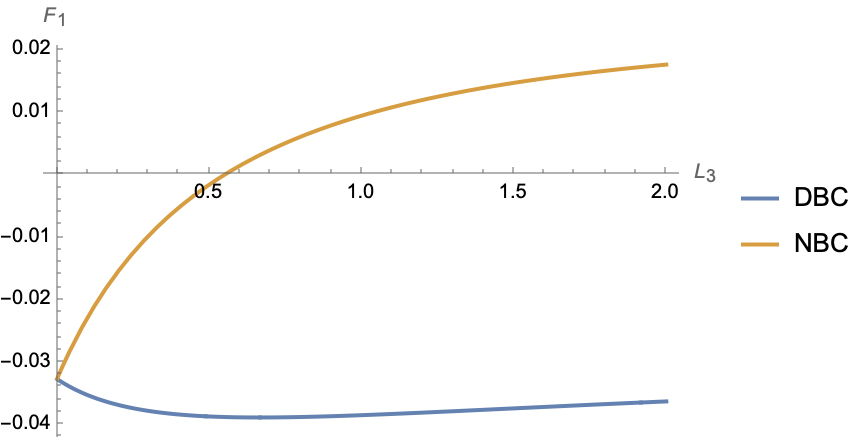}  \ \ \  \includegraphics[width=7cm]{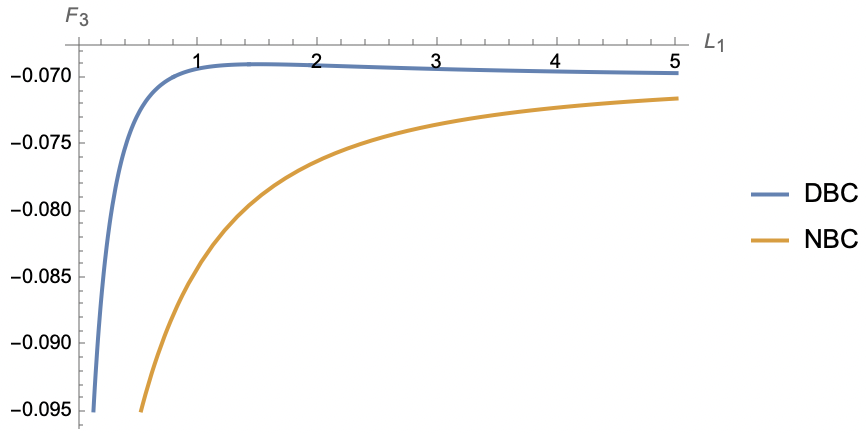}  
\caption{Casimir force for the network (Fig. \ref{Tree} (right)) with  $L_1=L_4$ and $L_2=L_3$. The left figure is for $F_1=F_4$ and the right figure is for $F_2=F_3$. The blue and yellow curves denote DBC and NBC, respectively. Interestingly, for the NBC, the force $F_1$ on the outside edge changes signs when increasing the size of loop $L_3$.  }
\label{forceIII}
\end{figure}

Finally, we conclude this subsection by presenting the Casimir forces $F^1(L_3)$ and $F^3(L_1)$ for the case where $L_4 = L_1$ and $L_2 = L_3$, shown in Fig. \ref{forceIII}. Notably, for the NBC, the force $F_1$ on the outside edge reverses direction as the loop size $L_3$ increases, similar to the behavior observed in the single-node network discussed in sect. 3.2.

\section{Generalizations to higher dimensions }

This section investigates the Casimir effect on the network in general dimensions. For simplicity, we focus on the network with n edges linked by one node in $1+(d-1)$ dimensional spacetime. Below, we use DBC as an example and then provide the results for NBC directly. Since the discussions are similar to those of sect. 3.1, we show only the key points below. 

The scalar obeying DBC (\ref{sect3: DBC}) on edge $E_i$ reads
\begin{eqnarray} \label{sect3.2: solution DBC}
\phi_i= a_i \sin\Big(k(x_i-L_i)\Big) e^{i \vec{k}_{\parallel} \cdot \vec{y}-i \omega t},
\end{eqnarray} 
where $\omega^2=k^2+k^2_{\parallel}$, and $\vec{y}$ and $\vec{k}_{\parallel}$ denote the transverse coordinates and wave vectors, respectively. Then the junction condition (\ref{sect2: junction condition}) on the node gives the spectrum
 \begin{eqnarray} \label{sect3: constraint NBC}
\Delta(k)=\sum_{i=1}^n \cot(k L_i)=0.
\end{eqnarray} 
It has the zeros $k_n$ fixed by $\Delta(k_n)=0$ and poles $q_m=m \pi/L_i$ with positive integer $m$ and $i=1,2,...,n$. In higher dimensions, the key formula to derive Casimir energy can be generalized to 
\begin{eqnarray} \label{sect3.2: key trick}
\frac{W}{A}&=&\int (\frac{dk_{\parallel}}{2\pi})^{d-2}\sum_n \frac{\omega_n}{2}\nonumber\\
&=&\int (\frac{dk_{\parallel}}{2\pi})^{d-2}\sum_m \frac{p_m}{2}+\int (\frac{dk_{\parallel}}{2\pi})^{d-2}\frac{1}{2\pi i}\oint \frac{\omega}{2} d \ln \Delta(\sqrt{\omega^2-k_{\parallel}^2}),
\end{eqnarray} 
where $A=\int dy^{d-1}$ is the transverse area, $\omega_n=\sqrt{k_n^2+k_{\parallel}^2}$ and $p_m=\sqrt{q_m^2+k_{\parallel}^2}$ with $k_n$ and $q_m=m \pi/L_i$ are zeros and poles of (\ref{sect3: constraint NBC}), respectively. For simplicity, we label $W/A$ by $\bar{W}$, and the first (second) term of (\ref{sect3.2: key trick}) by $\bar{W}_1$($\bar{W}_2$).  Note that $\bar{W}_1$ is just the usual Casimir energy between parallel planes
\begin{eqnarray} \label{sect3.2: W1}
\bar{W}_1&=& \lim_{s\to 0}\sum_{i=1}^n \frac{S(d-3)}{(2 \pi )^{d-2}}\int_0^{\infty} dk_{\parallel} k_{\parallel}^{d-3} \sum_{m=1}^{\infty} \frac{1}{2}(\frac{m^2\pi^2}{L_i^2}+k_{\parallel}^2)^{\frac{1}{2}+s} \nonumber\\
&=&-2^{-d} \pi ^{\frac{d-1}{2}} \zeta (1-d) \Gamma \left(\frac{1}{2}-\frac{d}{2}\right)\sum_{i=1}^n \frac{1}{L_i^{d-1}},
\end{eqnarray}
where $S(d-3)=2 \pi ^{\frac{d-2}{2}}/\Gamma \left(\frac{d-2}{2}\right)$ is the area of unite $(d-3)$-dimensional sphere. Following the approach of sect.3.1, we derive 
\begin{eqnarray} \label{sect3.2: W2}
\bar{W}_2&=&2^{-d} \pi ^{1-d}\int_{-\infty}^{\infty} dk_{\parallel}^{d-2} \int_{-\infty}^{\infty}d\omega  \frac{w^2 \left(\sum_{i=1}^nL_i \text{csch}^2\left(L_i \sqrt{k_{\parallel}^2+w^2}\right)\right)}{\sqrt{k_{\parallel}^2+w^2} \sum_{j=1}^n\coth \left(L_j \sqrt{k_{\parallel}^2+w^2}\right)},\nonumber\\
&=&\frac{2^{-d} \pi ^{\frac{1}{2}-\frac{d}{2}}}{\Gamma \left(\frac{d+1}{2}\right)} \int_0^{\infty} dr \frac{r^{d-1} \sum_{i=1}^nL_i \text{csch}^2\left(L_i  r\right)}{ \sum_{j=1}^n\coth \left(L_j r\right)},
\end{eqnarray}
where we have used $r^2=k_{\parallel}^2+\omega^2$. Thus, the total Casimir energy per area reads
\begin{eqnarray} \label{sect3.2: W DBC}
\frac{W_{\text{DBC}}}{A}&=&-2^{-d} \pi ^{\frac{d-1}{2}} \zeta (1-d) \Gamma \left(\frac{1}{2}-\frac{d}{2}\right)\sum_{i=1}^n \frac{1}{L_i^{d-1}}\nonumber\\
&+&\frac{2^{-d} \pi ^{\frac{1}{2}-\frac{d}{2}}}{\Gamma \left(\frac{d+1}{2}\right)} \int_0^{\infty} d\omega \frac{\omega^{d-1} \sum_{i=1}^nL_i \text{csch}^2\left(L_i  \omega\right)}{ \sum_{j=1}^n\coth \left(L_j \omega\right)}.
\end{eqnarray}
From the above Casimir energy, we can derive the pressure on edge $E_i$ as $P^i=-(\partial W/\partial L_i)/ A$. Noting that 
\begin{eqnarray} \label{sect3.2: DBC trick}
 \int_0^{\infty} d\omega \frac{\omega^{d-1} \sum_{i=1}^nL_i \text{csch}^2\left(L_i  \omega\right)}{ \sum_{j=1}^n\coth \left(L_j \omega\right)}&=&-\int_0^{\infty}\omega^{d-1} d\log\Big(\frac{ \sum_{j=1}^n\coth \left(L_j \omega \right)}{n} \Big)\nonumber\\
 &=&(d-1) \int_0^{\infty} d\omega \omega^{d-2} \log\Big(\frac{ \sum_{j=1}^n\coth \left(L_j \omega \right)}{n} \Big),
\end{eqnarray}
we derive 
\begin{eqnarray} \label{sect3.2: Pi DBC}
P^i_{\text{DBC}}&=&-(d-1)2^{-d} \pi ^{\frac{d-1}{2}} \zeta (1-d) \Gamma \left(\frac{1}{2}-\frac{d}{2}\right) \frac{1}{L_i^{d}}\nonumber\\
&+&(d-1)\frac{2^{-d} \pi ^{\frac{1}{2}-\frac{d}{2}}}{\Gamma \left(\frac{d+1}{2}\right)} \int_0^{\infty} d\omega \frac{\omega^{d-1} \text{csch}^2\left(L_i  \omega\right)}{ \sum_{j=1}^n\coth \left(L_j \omega\right)}.
\end{eqnarray}
Note that (\ref{sect3.2: W DBC}) and (\ref{sect3.2: Pi DBC}) obeys the conformal relation $(d-1)T_{tt}=T_{xx}$ for CFT in $d$ dimensions 
\begin{eqnarray} \label{sect3.2: conformal relation}
(d-1)\frac{W}{A}=\sum_{i=1}^n P^i L_i,
\end{eqnarray}
which provides strong support for our results. By applying the same approach, we derive for NBC
\begin{eqnarray} \label{sect3.2: W NBC}
\frac{W_{\text{NBC}}}{A}&=&-2^{-d} \pi ^{\frac{d}{2}-\frac{1}{2}} \zeta \left(1-d,\frac{1}{2}\right) \Gamma \left(\frac{1}{2}-\frac{d}{2}\right)\sum_{i=1}^n \frac{1}{L_i^{d-1}}\nonumber\\
&&-\frac{2^{-d} \pi ^{\frac{1}{2}-\frac{d}{2}}}{\Gamma \left(\frac{d+1}{2}\right)} \int_0^{\infty} d\omega \frac{\omega ^{d-1} \sum_{i}^n L_i \text{sech}^2\left(\omega  L_i\right)}{\sum _{j=1}^n \tanh \left(\omega  L_j\right)},
\end{eqnarray}
and 
\begin{eqnarray} \label{sect3.2: Pi NBC}
P^i_{\text{NBC}}&=&-(d-1)2^{-d} \pi ^{\frac{d}{2}-\frac{1}{2}} \zeta \left(1-d,\frac{1}{2}\right) \Gamma \left(\frac{1}{2}-\frac{d}{2}\right) \frac{1}{L_i^{d}}\nonumber\\
&&-(d-1)\frac{2^{-d} \pi ^{\frac{1}{2}-\frac{d}{2}}}{\Gamma \left(\frac{d+1}{2}\right)} \int_0^{\infty} d\omega \frac{\omega ^{d-1} \text{sech}^2\left(\omega  L_i\right)}{\sum _{j=1}^n \tanh \left(\omega  L_j\right)}.
\end{eqnarray}
It is easy to see that the Casimir energy (\ref{sect3.2: W NBC}) and pressure (\ref{sect3.2: Pi NBC}) for NBC also obeys the conformal relation (\ref{sect3.2: conformal relation}). We verify that the Casimir energy (\ref{sect3.2: W DBC},\ref{sect3.2: W NBC}) and Casimir pressure (\ref{sect3.2: Pi DBC},\ref{sect3.2: Pi NBC}) agree with these of sect.3 for $d=2$, which is a test of our results.

\begin{figure}[t]
\centering
\includegraphics[width=7cm]{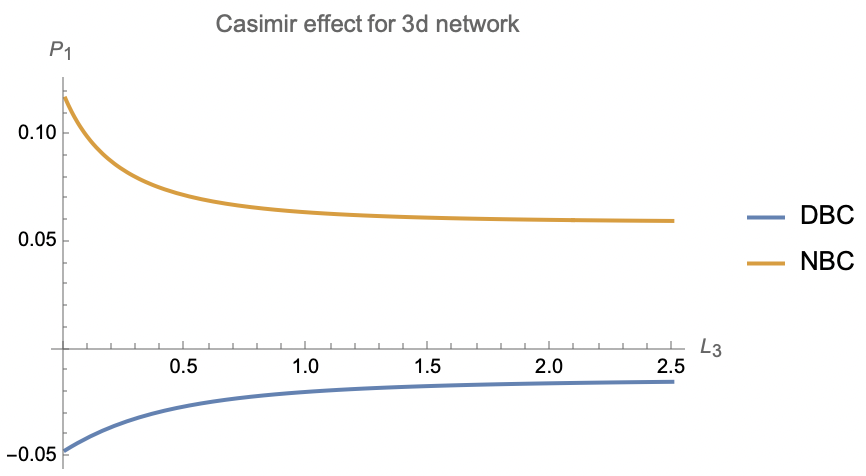} \includegraphics[width=7cm]{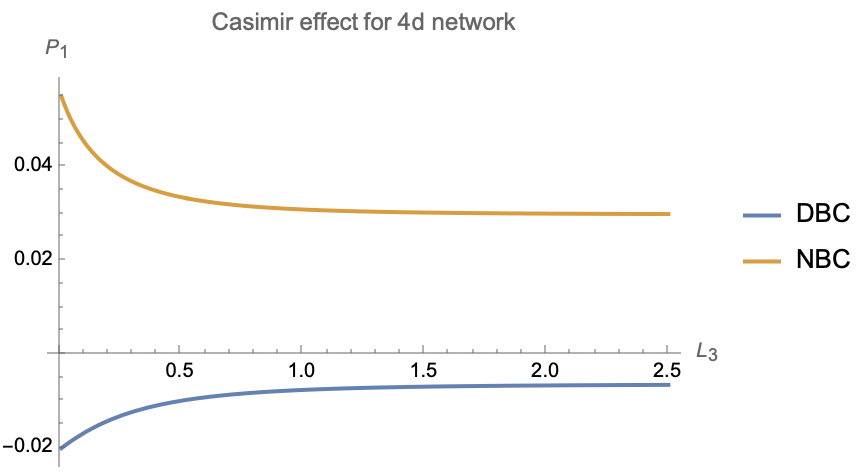} 
\caption{ Casimir effect for 3d (left) and 4d (right) networks with $L_1=1, L_2=L_3$. It shows the Casimir pressure on edge $E_1$ varies with the length of edge $E_3$ and approaches a constant value in the large $L_3$ limit. Unlike the 2d case, the Casimir force does not change signs. Here by ``3d", we means $(1+2)$-dimensional spacetime. }
\label{3d4dnetwork}
\end{figure}

To conclude this subsection, we illustrate $P_1(L_3)$ for 3d and 4d networks with three edges connected by one node in Fig. \ref{3d4dnetwork}. Here, ``3d" refers to a $(1+2)$-dimensional spacetime. We use the example where $L_1 = 1$ and $L_2 = L_3$. In Fig. \ref{3d4dnetwork}, the Casimir pressure on edge $E_1$ varies with the length of edge $E_3$ for both DBC and NBC, approaching a constant value as $L_3$ gets large. Similar to the 2d case, the Casimir force is repulsive for NBC. Unlike the 2d case, the Casimir pressure does not change signs in 3d and 4d networks.

\section{Conclusions and Discussions}

This paper examines the Casimir effect in networks, utilizing a free massless scalar field as an illustrative example. We derive the junction condition at network nodes, ensuring it aligns with current conservation—meaning the total currents entering a node equal zero. We discover that the Casimir force on one edge can be switched from attractive to repulsive by adjusting the lengths of other edges, assuming we apply Neumann boundary conditions (NBC) on the outer edges. It provides a practical means of controlling the Casimir effect, with potential applications in nanotechnology. Additionally, we find that the Casimir amplitude for any edge is smaller than that of a strip of the same length. This occurs because waves can move between edges, causing the node to act like a semi-transparent boundary, which reduces the Casimir amplitude. We first examine the simplest network in 1+1 dimensions and later extend our findings to more complex networks and higher dimensions. In higher dimensions, we also observe a repulsive Casimir force for NBC. However, unlike in lower dimensions, the force does not change sign when we adjust the edge lengths; instead, there is a minimum value for the Casimir force on one edge when the other edges are of infinite length.

Many interesting problems are worthy of exploration. Let us list just a few of them. 

\begin{itemize}

 \item This paper focuses on scalar fields and planar networks. It would be interesting to extend the study to include different types of fields, like Dirac fermions and Maxwell fields, as well as 3D networks. Considering the temperature effect \cite{Bordag:2001qi, Chen:2024bis} is also interesting. 
  
 \item Exploring other quantum effects, such as entanglement and tunneling in networks, could provide valuable insights into these phenomena. Additionally, solving the Schrödinger equation in the network is also intriguing, especially since the junction condition for wave functions may alter energy levels and introduce new phenomena.
 
  \item Recently, one of the authors proposed a fundamental bound of the Casimir effect \cite{Miao:2024gcq, Miao:2025utb}. As we discussed above, the Casimir effect in the network becomes smaller than that of a strip. It is intriguing to explore the bounds of the Casimir effect in a network with $n$ nodes and $p$ edges. 
   
 \item Finally, using realistic 3D devices to simulate 1D networks experimentally could help validate our predictions. The Casimir effect in networks made of thin tubes is another interesting area to investigate. 

 \end{itemize}

\section*{Acknowledgements}
% Zhao is supported by Guangdong Basic and Applied Basic Research Foundation (No.2021A1515010276) and National Natural Science Foundation of China (NSFC) grant (No.11904422). 
We thank S.N Pang, L. Li and Y. Guo, for helpful discussions. Miao acknowledges the supports from NSFC (No.12275366).

\section*{Appendix}

In this appendix, we take the Green-function method to study the Casimir effect on the network. For simplicity, we focus on the conformal scalar in the $d$-dimensional Euclidean spacetime $M=R^{d-1}\times N^1$. Here $R^{d-1}$ is the $(d-1)$-dimensional flat space, and $N^1$ denotes the one-dimensional network with n edges linked by one node (Fig. \ref{network}). The coordinates on the edge $E_i$ are $(x_i, y^a)$ with $i=1,2,...,n$ and $a=1,2,..,d-1$. We have $0\le x_i\le L_i$ with the node at $x_i=0$. We take $x_{\mu}$ to label the coordinates on the whole space $M=R^{d-1}\times N^1$.  

Let us start with the heat kernel obeying EOM
 \begin{eqnarray}\label{app: EOM heat kernel}
(\partial_s -\Box) K(s,x_{\mu},x'_{\mu})=0,
\end{eqnarray}
and the limit
 \begin{eqnarray}\label{app: BC heat kernel}
\lim_{s\to 0} K(s,x_{\mu},x'_{\mu})=\delta^{(d)}(x_{\mu}-x'_{\mu}). 
\end{eqnarray}
Here $\Box=\partial_{\mu}\partial_{\mu}$ is the D'Alembert operator, and $\delta^{(d)}(x_{\mu}-x'_{\mu})$ is the $d$-dimensional delta function. From the heat kernel, we can derive the Green function as \cite{Vassilevich:2003xt}
 \begin{eqnarray}\label{app: Greenfunction}
G(x_{\mu},x'_{\mu})=\int_0^{\infty} ds K(s,x_{\mu},x'_{\mu}),
\end{eqnarray}
and then derive the renormalized stress tensor by \cite{Miao:2024ddp}
\begin{eqnarray}\label{app: Tij}
\langle T_{\mu\nu} \rangle=\lim_{x'_{\mu}\to x_{\mu}} \left[  (1-2\xi) \partial_{\mu}\partial_{\nu'}-2\xi \partial_{\mu} \partial_{\nu} +(2\xi -\frac{1}{2})\delta_{\mu\nu} \partial_{\alpha} \partial_{\alpha'}  \right] \hat{G}(x_{\mu},x'_{\mu}),
\end{eqnarray}
where $\xi=\frac{d-2}{4(d-1)}$ and $\hat{G}=G-G_0$ with $G_0$ the Green function in the free space.

For the space $R^{d-1}\times N^1$, the heat kernel is given by 
\begin{eqnarray}\label{app: heat kernel 1}
K=K_{R^{d-1}} K_{N^1},
\end{eqnarray}
where \cite{Vassilevich:2003xt}
\begin{eqnarray}\label{app: heat kernel R}
K_{R^{d-1}}=\frac{1}{(4\pi s)^{\frac{d-1}{2}}} \exp(-\frac{\sum_{a=1}^{d-1} (y_{a}-y'_{a})^2}{4s}),
\end{eqnarray}
and $K_{N^1}$ can be obtained by 
\begin{eqnarray}\label{app: heat kernel N}
K_{N^1}=\sum_m e^{-\lambda_m s}\varphi_m(x_i) \varphi_m(x'_j),
\end{eqnarray}
where $\lambda_m$ and $\varphi_m(x_i)$ are the eigenvalues and eigenfunctions of D'Alembert operator on $N^1$
\begin{eqnarray}\label{app: eigenfunctions}
\partial^2_{x_i}\varphi_m(x_i)=-\lambda_m \varphi_m(x_i). 
\end{eqnarray}
Note that $\varphi_m(x_i)$ obeys the junction condition (\ref{sect2: junction condition}) on the node and either DBC (\ref{sect3: DBC}) or NBC (\ref{sect3: NBC}) on the boundaries. As a result, we have from sect. 3 that 
\begin{eqnarray}\label{app: eigenfunctions 1}
\varphi_m(x_i)=a_{m (i)}\begin{cases}
 \sin\Big( k_{m} (x_i-L_i) \Big),\ \ \ \ \ \text{for DBC} ,\\
\cos\Big( k_{m}  (x_i-L_i) \Big),\  \ \ \ \ \text{for NBC},
\end{cases} 
\end{eqnarray}
and $\lambda_m=k_m^2$ with $k_m$ obeying the spectrum $\Delta(k_m)=0$
\begin{eqnarray}\label{app: spectrum}
\Delta(k)=\begin{cases}
\sum_{i=1}^n \cot(k L_i),\ \ \ \ \ \text{for DBC} ,\\
\sum_{i=1}^n \tan(k L_i),\  \ \ \ \ \text{for NBC}.
\end{cases} 
\end{eqnarray}
We normalize the eigenfunctions as
\begin{eqnarray}\label{app: eigenfunctions normalize}
\sum_{i=1}^n\int_0^{L_i} \varphi_m(x_i)\varphi_n(x_i) dx_i=\delta_{mn},
\end{eqnarray}
which fixes the coefficients
\begin{eqnarray}\label{app: coefficients}
a_{m (i)}=\begin{cases}
\frac{\sqrt{2} \csc \left(L_i k_m\right)}{\sqrt{\sum _{j=1}^n L_j \csc ^2\left(L_j k_m\right)}},\ \ \ \ \ \text{for DBC} ,\\
\frac{\sqrt{2} \sec \left(L_i k_m\right)}{\sqrt{\sum _{j=1}^n L_j \sec ^2\left(L_j k_m\right)}},\  \ \ \ \ \text{for NBC}.
\end{cases} 
\end{eqnarray}
From (\ref{app: heat kernel 1},\ref{app: heat kernel R},\ref{app: heat kernel N},\ref{app: eigenfunctions 1},\ref{app: coefficients}), we obtain the heat kernel and then can derive the renormalized stress tensor from (\ref{app: Greenfunction},\ref{app: Tij}). For the conformal scalar, we have $T_{xx}=-(d-1) T_{y_ay_a}$ so that $\langle T^{\mu}_{\ \mu} \rangle=0$. Without loss of generality, we focus on the pressure $T_{xx}$ below. 

From (\ref{app: Tij}), we obtain the pressure on the edge $E_i$
\begin{eqnarray}\label{app: Pi}
P^i=\langle T_{xx} \rangle|_{E_i}=2^{1-d} \pi ^{\frac{1}{2}-\frac{d}{2}} \Gamma \left(\frac{3}{2}-\frac{d}{2}\right) 
\begin{cases}
\sum_{m}  \frac{k_m^{d-1} \csc^2\left(L_i k_m\right)}{\sum _{j=1}^n L_j \csc ^2\left(L_j k_m\right)},\ \ \ \ \ \text{for DBC} ,\\
\sum_{m}  \frac{k_m^{d-1} \sec^2\left(L_i k_m\right)}{\sum _{j=1}^n L_j \sec ^2\left(L_j k_m\right)},\  \ \ \ \ \text{for NBC},
\end{cases} 
\end{eqnarray}
 where $k_m$ are zeros of (\ref{app: spectrum}). In the above derivations, we first take the limit $x'_{\mu}\to x_{\mu}$, and then do the integral of $s$ for suitable dimension $d$. Finally, we analytically continue the results to general even $d$ and get the above result. Note that, for odd $d$, (\ref{app: Pi}) is divergent and we need further regularizations. For simplicity, we focus on even $d$ below. 
 
We take the following formula to compute (\ref{app: Pi})
\begin{eqnarray} \label{app: key trick}
P^i=\sum_n f(\omega_n)=\sum_m f(p_m)+\frac{1}{2\pi i}\oint f(\omega) d \ln \Delta(\omega),
\end{eqnarray} 
where $\omega_n$ and $p_m$ are zeros and poles of $\Delta(k)$ (\ref{app: spectrum}).  The poles of (\ref{app: spectrum}) are given by
\begin{eqnarray}\label{app: poles}
p_{m}=\begin{cases}
\frac{m \pi}{L_i},\ \ \ \ \ \ \ \ \ \  \text{for DBC} ,\\
\frac{(\frac{1}{2}+m) \pi}{L_i},\  \ \ \ \ \text{for NBC}.
\end{cases} 
\end{eqnarray}
Thus, the pole contributions to the pressure (\ref{app: key trick}) is 
\begin{eqnarray} \label{app: pole contribution}
\sum_m f(p_m)&=&2^{1-d} \pi ^{\frac{1}{2}-\frac{d}{2}} \Gamma \left(\frac{3}{2}-\frac{d}{2}\right) \begin{cases}
\sum_{m}\frac{(m \pi)^{d-1} }{ L_i^d},\ \ \ \ \ \ \ \text{for DBC} ,\\
\sum_{m} \frac{(\frac{1}{2}\pi+m \pi)^{d-1} }{ L_i^d},\  \ \ \text{for NBC},
\end{cases} \nonumber\\
&=&2^{1-d} \pi ^{\frac{d}{2}-\frac{1}{2}} \Gamma \left(\frac{3}{2}-\frac{d}{2}\right)\frac{1}{L_i^d} \begin{cases}
\zeta(1-d) ,\ \ \ \ \ \ \ \text{for DBC} ,\\
\zeta(1-d,\frac{1}{2}),\  \ \ \ \text{for NBC},
\end{cases}
\end{eqnarray} 
which agrees with the first line of (\ref{sect3.2: Pi DBC},\ref{sect3.2: Pi NBC}). The integral contribution to the pressure (\ref{app: key trick}) is
\begin{eqnarray} \label{app: integral contribution}
\frac{1}{2\pi i}\oint f(\omega) d \ln \Delta(\omega)=(-1)^{\frac{d}{2}} 2^{1-d} \pi ^{-\frac{d}{2}-\frac{1}{2}} \Gamma \left(\frac{3}{2}-\frac{d}{2}\right)\begin{cases}
-\int_0^{\infty} d\omega \frac{\omega ^{d-1} \text{csch}^2\left(\omega  L_i\right)}{\sum _{j=1}^n \coth \left(\omega  L_j\right)},\  \text{for DBC} ,\\
\int_0^{\infty} d\omega\frac{\omega ^{d-1} \text{sech}^2\left(\omega  L_i\right)}{\sum _{j=1}^n \tanh \left(\omega  L_j\right)},\  \ \ \ \text{for NBC}.
\end{cases} 
\end{eqnarray} 
Now we obtain the total pressure for DBC
\begin{eqnarray}\label{app: Pi DBC}
P^i_{\text{DBC}}&=&2^{1-d} \pi ^{\frac{d}{2}-\frac{1}{2}} \Gamma \left(\frac{3}{2}-\frac{d}{2}\right)\zeta(1-d)\sum_{i=1}^n\frac{1}{L_i^d} \nonumber\\
&&-(-1)^{\frac{d}{2}} 2^{1-d} \pi ^{-\frac{d}{2}-\frac{1}{2}} \Gamma \left(\frac{3}{2}-\frac{d}{2}\right)\int_0^{\infty} d\omega \frac{\omega ^{d-1} \text{csch}^2\left(\omega  L_i\right)}{\sum _{j=1}^n \coth \left(\omega  L_j\right)},
\end{eqnarray}
and for NBC
\begin{eqnarray}\label{app: Pi NBC}
P^i_{\text{NBC}}&=&2^{1-d} \pi ^{\frac{d}{2}-\frac{1}{2}} \Gamma \left(\frac{3}{2}-\frac{d}{2}\right)\zeta(1-d,\frac{1}{2})\sum_{i=1}^n\frac{1}{L_i^d} \nonumber\\
&&+(-1)^{\frac{d}{2}} 2^{1-d} \pi ^{-\frac{d}{2}-\frac{1}{2}} \Gamma \left(\frac{3}{2}-\frac{d}{2}\right)\int_0^{\infty} d\omega \frac{\omega ^{d-1} \text{sech}^2\left(\omega  L_i\right)}{\sum _{j=1}^n \tanh \left(\omega  L_j\right)}.
\end{eqnarray}
Recall that the above formulas work only for even $d$. For $d=2$ and $n=3$, the pressures (\ref{app: Pi DBC},\ref{app: Pi NBC}) reduces to (\ref{sect3: Casimir force DBC},\ref{sect3: Casimir force NBC}) of sect. 3.1. For general even $d$,  the pressures (\ref{app: Pi DBC},\ref{app: Pi NBC}) also agree with those of sect. 3.2. We now complete the discussion of the network Casimir effect by applying the Green-function method.

\end{document}